\documentstyle[proceedings]{crckapb}

\begin{opening}
\title{A new classification and a  two-parameter \protect \\
          unification of BL Lacertae objects }

\author{Markos Georganopoulos and Alan P. Marscher}
\institute{Department of Astronomy, Boston University\\
            725 Commonwealth Avenue, Boston, MA 02215, USA}

\end{opening}

\runningtitle{The accelerating inner jet}

\begin{document}

\begin{abstract}
 A new continuous classification system for BL Lacertae (BL Lac) objects is 
proposed. 
The peak frequency $\nu_{p}$ of the synchrotron component in the 
 $\nu L_{\nu}$ spectrum is used as a classifying 
parameter. The application of the 
accelerating inner jet model to  complete BL Lac samples 
suggests that the range of the observed properties may be explained by 
adopting 
a description based on two parameters: the jet angle to the line of sight
 $\Theta$, and the relativistic electron kinetic luminosity 
$\Lambda_{e}$. 
\end{abstract}

\section{Classification of BL Lacertae objects}

\begin{figure}
\vspace{10cm}
\includegraphics{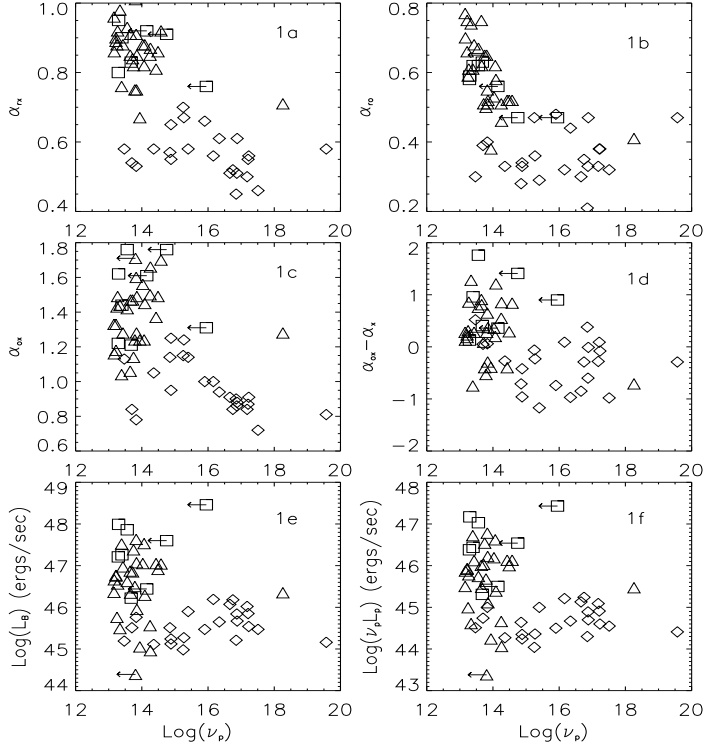}
\caption{The (a) radio to X-ray $(\alpha_{rx})$, (b) radio to optical 
($\alpha_{ro}$), (c)  optical to X-ray ($\alpha_{ox}$)
 broadband spectral indices, 
(d) the X-ray concavity index $\alpha_{ox}-\alpha_{x}$, 
(e) the bolometric synchrotron luminosity $L_{B}$, 
and (f) the peak luminosity $\nu_{p} L_{p}$ 
as a function of the peak frequency $\nu_{p}$ 
for the XBL (diamonds),
 RBL (triangles), and FSRQ (squares) samples of SMU96. 
Arrows indicate objects with only an estimated upper limit to  
 the peak frequency $\nu_{p}$.}
\label{data}
\end{figure}

BL Lac objects are discovered mostly through X-ray or 
radio surveys. The existing complete samples therefore are either X-ray 
 \cite{morris91} or radio flux-limited samples 
\cite{stickel91} and BL Lac objects are classified as XBL (X-ray selected)
 or RBL (radio selected) depending  on the method of discovery.
The $\nu L_{\nu}$ spectra of RBLs  peak 
somewhere in the far-infrared and have a higher peak 
luminosity than the 
XBLs that usually peak in the UV - soft X-ray
range (\citeauthor{sambruna96} \citeyear{sambruna96}, hereafter SMU96). 
According to a recently  proposed scheme 
\cite{padovani95}, BL Lac objects are classified as high peak frequency 
 BL Lac objects 
(HBL) or low peak frequency   BL Lac objects (LBL), depending on the value
of the X-ray to radio-flux ratio.
Most of the HBLs are XBLs and most of the LBLs are RBLs. 
Although this classification scheme reflects more closely the  observed 
properties of the BL Lac objects, it inherits the dichotomization 
that was imposed by the two different discovery methods. 
If the two methods of discovery force
us to observe the two extremes of a continuous distribution, thereby 
introducing  
 a  selection induced bimodality, the question that  arises 
is whether  there is an observed quantity
that varies in a continuous fashion, has a well defined physical significance,
and can be used as a classifying parameter.

\section{A new classification system}

Recently SMU96 examined the multifrequency spectral properties of three
complete samples of blazars, the \sl Einstein \rm Extended Medium Sensitivity 
Survey sample of XBLs \cite{morris91}, the $1 $ Jy 
sample of RBLs \cite{stickel91}, and a small complete sample of FSRQ from 
the S5 survey \cite{brunner94}. If we consider the BL Lac  
samples collectively (fig. \ref{data}) we notice that in all six diagrams there
is an upper envelope separating the populated area from a 
 well defined zone of avoidance. For a given peak frequency $\nu_{p}$ there
 is a permitted range of  
luminosities and spectral indices. As $\nu_{p}$ increases the maximum observed
luminosity decreases, the steepest observed spectral index flattens, and the
maximum observed concavity index $\alpha_{ox}-\alpha_{x}$ decreases. 
The flattest spectral indices  and the most negative concavity
indices do not seem to be very sensitive functions of  $\nu_{p}$. 

\it We propose a  continuous classification system for BL Lac objects,
 based on the
peak frequency  $\nu_{p}$ of the  $\nu L_{\nu}$ synchrotron 
luminosity distribution\rm.  A BL Lac
object is classified according to its peak frequency $\nu_{p}$. For example, 
a BL Lac object peaking at $log(\nu_{p})\cong14$ will be classified as a 
\it BL14 \rm 
object. 

The peak frequency of a BL Lacertae object is  a very 
important parameter from a physical point of view. Energetically, it is the 
most important frequency for the observed synchrotron radiation. It is
 closely related to the peak of the inverse Compton emission in both the
Synchrotron-self Compton (SSC) \cite{bloom96} and the 
external Compton (EC) models \cite{sikora94}. It is linked  to 
the energetics and geometry of the synchrotron source 
and to the angle formed between the observer and the  plasma  
bulk velocity, if  the source is moving relativistically 
(\citeauthor{georgano96} \citeyear{georgano96}).
 As SMU96 pointed out the accurate determination of $\nu_{p}$ requires 
contemporaneous flux measurements in different wavelength regimes. 
The contribution of new telescopes such as ISO towards this goal can be 
particularly important.

\section{The $\Theta-\Lambda_{e}$ scheme} 

Orientation and physical differences have been evoked to explain the range 
of the observed properties of BL Lac objects  
 (SMU96 ; \citeauthor{georgano96}  \citeyear{georgano96}). 
Under the proposed taxonomical scheme the problem of the XBL-RBL differences 
can be restated as follows: \it What is changing  as we continuously shift
 from a BL13 to a BL17 object? \rm 

Recently, 
\citeauthor{georgano96} \shortcite{georgano96}
 used the accelerating inner jet model to argue that,
 although  the mean properties
of  XBL and  RBL samples  can be explained under the orientation-determined
scenario, the sources  seem to be additionally characterized by a range of
physical parameters. It seems therefore necessary  to invoke the scaling of 
an appropriate
physical quantity that would   preserve the observed self-similarity of the 
relativistic jets from the galactic to the extragalactic scale. 
An observed physical parameter that is known to vary
over several decades in relativistic jets and over almost  three decades in BL Lac
objects is the synchrotron peak luminosity $L_{p}$. A plausible way
to explain this range of $L_{p}$  is to associate it 
with an intrinsic range of the relativistic electron kinetic luminosity 
$\Lambda_{e}$. In a scale invariant
description of the jet, the luminosity relates to the jet radius according to 
the relation $r\propto\Lambda_{e}^{1/2}$ \cite{georgano97}. 

\begin{figure}
\vspace{6.cm}
\includegraphics{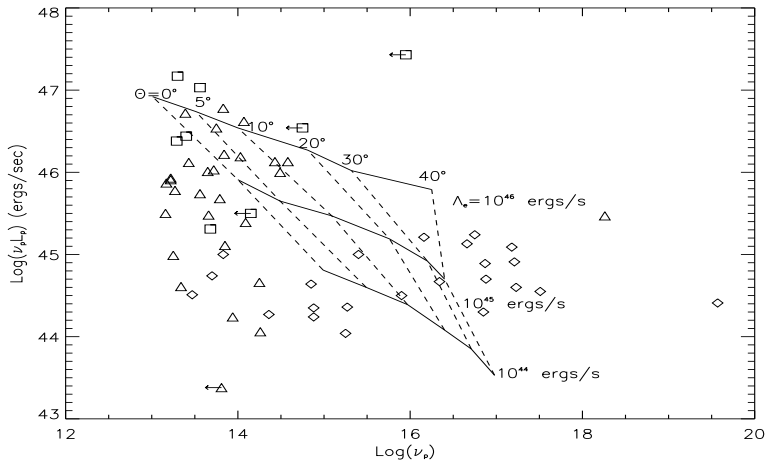}
\caption{Peak luminosity $\nu_{p}L_{p}$ versus peak frequency $\nu_{p}$. 
The data points are as in fig. \ref{data}. The solid lines correspond to the 
migration of the model prediction as the angle $\Theta$ changes from $0^{o}$
 to $40^{o}$ for a set of  electron kinetic luminosity $\Lambda_{e}$. 
The dashed lines are lines of constant angle $\Theta$.}
\label{model}
\end{figure}

Preliminary results (fig. \ref{model}) show that a combination of a range of 
electron  kinetic luminosities $\Lambda_{e}$ and 
jet orientation angles $\Theta$ covers a major part of the observed 
$\nu_{p}-\L_{p}\nu_{p}$ space. It is very encouraging that this can be
 achieved only  through 
variation of aspect and jet power.
 We are continuing  our work   by comparing
 the predicted and 
observed properties of BL Lac objects as well as by  incorporating 
the flat spectrum radio quasars in the 
above scheme.

\end{document}